\documentclass[12pt]{article}
\newcommand{\bb}{\bibitem}
\newcommand{\cc}{\cite}

\newcommand{\sss}{\sigma}

\newcommand{\Om}{\Omega}
\newcommand{\om}{\omega}
\newcommand{\lt}{\left}
\newcommand{\rt}{\right}
\newcommand{\lll}{\lambda}
\newcommand{\F}{{\cal F}}

\newcommand{\Ss}{\hat S^2_}
\newcommand{\R}{\bar R}
\newcommand{\A}{\hat A_a}
\newcommand{\B}{\hat B_b}
\newcommand{\Aa}{\hat A_{a'}}
\newcommand{\Bb}{\hat B_{b'}}

\newcommand{\bea}{\begin{eqnarray} \label}
\newcommand{\eeq}{\end{equation}}
\newcommand{\beq}{\begin{equation} \label}
\newcommand{\eea}{\end{eqnarray}}
\newcommand{\nn}{\\ \nonumber}
\newcommand{\rr}[1]{(\ref{#1})}

\author {D.A.Slavnov}

\title {Quantum measurements and Kolmogorovian probability theory}

   \date {}

\begin {document} 

  \maketitle

\begin{center} {\it  Department of Physics, Moscow State
University,\\  Moscow 119992, Russia. E- mail:
slavnov@goa.bog.msu.ru } \end{center}

 \begin {abstract}

We establish connections between the requirement of measurability
of a probability space and the principle of complimentarity in
quantum mechanics. It is shown that measurability of a probability
space implies the dependence of results of quantum measurement not
only on the properties of a  quantum object under consideration,
but also on the classical characteristics of the measuring device
which is used. We show that if one takes into account the
requirement of measurability in a quantum case, the Bell
inequality does not follow from the hypothesis about the existence
of an objective reality.
\end {abstract}

\section {Introduction}

In the review \cc{hom} by Home and  Whitaker, devoted to
interpretation of quantum mechanics one can find the statement: "The
fundamental difficulty in interpretation of quantum theory is that
it provides in general  only  probabilities of obtaining given
results. Thus much of any discussion of quantum theory must depend
on what one means by probability --- how one defines or interprets
the term".

In this review a lot of place is occupied by the interpretation of
concept "probability"  but  Kolmogorovian approach~\cc {kol}
is mentioned only casually. At the same time namely Kolmogorovian
probability theory is the most consecutive and mathematically
strict. Besides, it most fully corresponds to that particular situation
which takes place in quantum mechanics in the author's opinion. In
present paper the application of Kolmogorovian probability theory
to the problem of quantum measurements will be described.

\section { Probability space}

Let us recollect original positions  of Kolmogorovian probability
theory (see, for example~\cc {nev}). The so-called probability
space $ (\Om, \F, P) $ lays in  the foundation of
probability-theoretic scheme. Here $ \Om $ is a set (space) of the
elementary events $ \om $. The elementary event is understood as a
possible result of  a single experiment. Besides the elementary
event the concept of "event" is  introduced. Each event $F $ is
identified with some subset of set $ \Om $. It is assumed that in
the experiment under consideration an event $F $ is carried out if
the result of experiment belongs to $F $ ($ \om\in F $).

Collections of subsets of the set $ \Om $ (including the set $ \Om
$ and the empty set $ \emptyset $) are alloted with the structure
of Boolean algebras. Algebraic operations are: intersection of
subsets, joining of them, and  complement with respect to $ \Om $.

The second ingredient  of a probability space is the so-called $ \sss
$-algebra $ \F $. It is some Boolean algebra, closed in respect of
denumerable number of operations of joining and intersection. The set
$ \Om $ in which the particular  $ \sss$-algebra $ \F $ is chosen,
refers to as measurable space. Further on the measurability will play a
key role.

Finally, the third ingredient  of probability space is a
probabilistic measure $P $. This is a mapping of algebra $ \F $
into the set of real numbers satisfying conditions: a) $0\leq P
(F) \leq 1 $ for all $F\in\F $, $P (\Om) =1 $; b) $P (\sum_j F_j)
= \sum_j P (F_j) $ for any denumerable collection of
nonintersecting subsets $F_j\in \F $. Let us pay attention that
the probabilistic measure is defined only for the events which are
included in the algebra $ \F $. For the elementary events the
probability, generally speaking, is not defined.

 The mapping $X $ of the set $ \Om $ in the expanded real straight line
  $ \R = [-\infty, + \infty] $ refers to as the real random quantity on
   $ \Om $. $$ X (\om) =X\in \R  $$
 It is supposed that the set $ \R $ is alloted by the property of
measurability.
In the set $ \R $ as a $ \sss $-algebra  we can take Boolean
algebra $ \F_R $, generated by semiopen intervals $ (x_i, x_j] $.
This is a $ \sss $-algebra which is obtained by an application of
algebraic operations to various intervals. We  designate by $
\{\om\in F_R \} $, $F_R\in \F_R $ the subset of elementary
events $ \om $ for which $X (\om) \in F_R $ . The subsets $F =
\{\om\in F_R \} $ form the $ \sss $-algebra $\F$ in the space $\Om $.

 \section {Quantum measurements}

We consider now the application of formulated main principles
of probability theory to a problem of quantum measurements. We
carry out our consideration starting from positions of "the objective
local theory", traced back to works of  Bell~\cc {bell1, bell2},
and also to the paper by Einstein, Podolsky, Rosen~\cc {epr}.

It is usually considered that "the objective local theory"
contradicts corollaries of the standard quantum mechanics. In present
paper we  try to show that such contradiction does not arise when
we use Kolmogorovian probability theory.

Let us suppose that there is a certain objective reality, which
determines possible result of any individual measurement.  We name
this objective reality by a physical state of a quantum object.
One should not mix this notion with what is usually denominated by
the term "state" in standard quantum mechanics. We shall use the
term "quantum state" in latter case.

It is possible to read  about a correlation between physical and
quantum states, and the mathematical concept corresponding to a
physical state in the papers~\cc{slav, slav1}.  Here we only note
that the quantum state is a certain equivalence class  of physical
states. This class has a potency of continual set. In the
probability theory we associate an elementary event with a
physical state. Correspondingly, we associate  the set of physical
states of a quantum object with the space $\Om $. Further, we need
to make this space measurable, i.e. to choose certain
$\sss$-algebra $\F$. Here, a peculiarity of quantum measurement,
which has the name "principle of complimentarity" in standard
quantum mechanics, has crucial importance.

As opposed to measurements in classical physics, in a quantum case
the measurements can be compatible and incompatible.
Correspondingly, in quantum mechanics the observable quantities
are subdivided into compatible (simultaneously measurable) and
incompatible (additional). In principle, measurements of
compatible observables can be carried out so that measurements of
one observable do not disturb the measurement of other observable. We
name the corresponding system of measuring devices in coordination with
collection of these compatible observables. As a matter of
principle  it  cannot be made for incompatible observables. As it
is told in the paper by Zeilinger~\cc{zeil}: "Quantum complementarity
then is simply an expression of the fact that in order to measure
quantities, we would have to use apparatuses which mutually
exclude each other".

Thus, we can organize each individual experiment only in such a
way that compatible observables are measured in it. The results of
measurement can be random. That is, observables correspond to the
real random quantities in probability theory.

The main purpose of typical quantum experiment is the finding of
probabilistic distributions of those or other observable
quantities. We can obtain such distribution for certain collection
of compatible observables on the application of certain measuring
device. From the point of view of probability theory we choose
certain $ \sss $-algebra $ \F $, choosing the certain measuring
device. For example, let us use the device intended for
measurement of momentum of a particle. Let us suppose that we can
ascertain by means of
this device  that the momentum of particle hits  an interval $
(p_i, p_j] $. For definiteness we have taken a semiopen interval
though it is not necessary. Hit of momentum of the particle in
this or that interval is the event for the measuring device, which we
use. These events are elements of certain $ \sss $-algebra. Thus, the
probability space $ (\Om, \F, P) $ is determined not only by the
explored quantum object (by  collection of its physical states)
but also by the measuring device which we  use.

Let us assume that we carry out some typical quantum experiment. We have
an ensemble of the quantum systems, belonging to a certain  {\it
quantum} state. For example, the particles have  spin 1/2 and the
spin projection on the $x $ axis equals 1/2. Let us
investigate the distribution of two incompatible observables (for
example,  the spin projections  on the directions forming angles
$\theta_1 $ and $ \theta_2 $ with regard to the $x $ axis). We
cannot measure both observables  in one experiment. Therefore, we
should carry out two groups of experiments which use
different measuring devices. "Different"  is  classically distinct.
In our concrete case the devices should be oriented by  various
manners in the space.

We can describe one group of experiments with the help of a
probability space $ (\Om, \F_1, P_1) $, another group with the help of $
(\Om, \F_2, P_2) $.  Although in both cases the space of
elementary events $ \Om $ is the same, the probability spaces are
different.  Certain $ \sss $-algebras $ \F_1 $ and $ \F_2 $ are
introduced in   these spaces to give them the property of
measurability.

Formally, only mathematically, in the space $ \Om $ we can
construct $ \sss $-algebra $\F_{12}$ which incorporates algebras
$\F_1 $ and $ \F_2 $. Such algebra refers to as generated by the
algebras $ \F_1 $ and $ \F_2 $. It contains besides the subsets
$F^{(1)}_i\in \F_1 $ and $F^{(2)}_j\in \F_1 $  also any possible
intersections and joins of the subsets $F^{(1)}_i\in \F_1 $ and
$F^{(2)}_j\in \F_2 $.

  However,  physically such $ \sss  $-algebra is unacceptable.
Really, the event $F_{ij} =F^{(1)}_i\cap F^{(2)}_j $ is that the
values of two incompatible observables belong to strictly defined
regions for one quantum object. It is  impossible to carry out
experiment which could pick out such event for a quantum system in
principle. Therefore, for such event there is no concept of
"probability". That is, there is no any
probabilistic measure corresponding to the subset $F_{ij}$, and the
$\sss$-algebra $ \F_{12} $ is not good for construction of a
probability space.

Here, basic characteristic of the application of probability
theory to quantum systems become apparent: not every
mathematically possible $ \sss $-algebra is physically allowable.

How  does the probability space  materialize in quantum
experiment? The definition of probability implies  repeated
realization of tests. We should carry out these tests in {\it
identical} conditions. It concerns both  the  object measured,
and  the measuring device. Evidently, we can supervise completely
a microstate for neither one nor other. Therefore, the term
"identical conditions" should be understood as some equivalence
classes of states of the quantum object and the measuring device.

We can supervise only classical characteristics of the physical
system. Therefore,  fixing some classical parameters of the
physical system, we fix a corresponding equivalence class. For the
object measured such fixing usually is the choice of the certain
quantum state. For example, for particles with  spin --- it is
selection of particles with the certain orientation of  spin. For
the measuring device we also should choose the certain classical
characteristic which fixes some equivalence class. For example, in
the measuring device the initial united beam of particles should
be split in several beams sufficiently well separated from
each other, corresponding to different values of the spin
projection onto a chosen direction.

Thus, in experiment the element of measurable space $ (\Om, \F) $
corresponds to the pair --- a quantum object (for example,
belonging to the certain quantum state) plus the certain type of
measuring device, which allows to fix events of the certain
form. Such device can pick out the events corresponding to
some set of compatible observable quantities. Therefore, in a
quantum case the measuring devices should be divided into various
types. Each type is coordinated with the separate collection  of
compatible observables.

Existence of various types of the measuring device leads  to one
more peculiarity of quantum measurements. The  same observable can
belong to two (or more) various collections   of compatible
observables. Such observable is compatible with all observables,
which are included in various collections. However, other
observables of different collections are not compatible among
themselves. For the measurement of the chosen observable we can
use various types of measuring device. It means that we can
carry out experiment in various conditions. There is no guarantee
that the result of measurement will not depend on these conditions.

Thus, the result of individual quantum measurement can depend not
only on the internal properties of a measured object (a physical
state), but also on the type of measuring device. In terms of
probability theory it is expressed as follows. For quantum system
the random quantity $X $ can be multiple-valued function of the
elementary event $ \om $.

  In the classical case all observables are compatible.
Correspondingly, all measuring devices belong to one type.
Therefore, the classical random quantity $X $ is the single-valued
function of $ \om $. Let us notice that if in the quantum case we
consider magnitude $X $ as a function not on space $ \Om $ but on
measurable space $ (\Om, \F) $, this function is single-valued.

This statement allows to look in a new fashion on the result
obtained in the paper by Kochen and Specker~\cc{ksp}, where the no-go
theorem is proved. The sense of this theorem can be expressed by
the statement that for a
particle with spin 1 there is no such internal characteristic
which uniquely predetermines the values of squares of the spin
projections on three mutually orthogonal directions.

In the standard quantum mechanics such three squares are described
by mutually commuting operators. Therefore, the corresponding
observables  $ (\Ss x, \Ss y, \Ss z) $ are compatible. The
observables $ (\Ss {x}, \Ss {y'}, \Ss {z'}) $ are also compatible.
Here, the $x, y', z' $ directions are orthogonal among themselves,
but the $y, z $ directions are not parallel to the $y', z' $
directions. The observables $ (\Ss y, \Ss z) $ are not compatible
with the observables $ (\Ss {y '}, \Ss {z '}) $. The devices
coordinated with the observables $ (\Ss x, \Ss y, \Ss z) $ and the
$ (\Ss {x}, \Ss {y '}, \Ss {z '}) $ belong to different types.
Therefore, these devices not necessarily should give the same
result for square of spin projection on the $x $ direction.

We cannot use simultaneously two types of measuring devices in one
experiment. Therefore, it is impossible to carry out direct
experiment for check of this statement. However, it is possible to
try to carry out indirect measurement.

For example, it is possible to organize such experiment as
follows. To take a physical system in singlet state which decays
in two massive particles with  spin 1. For one of particles to
measure $ \Ss x $ with the help of the device coordinated with the
observables $ (\Ss x, \Ss y, \Ss z) $, and for the other to
measure $ \Ss x $ with the help of the device coordinated with the
observables $ (\Ss {x}, \Ss {y '}, \Ss {z '}) $. After that it is
necessary to compare the results which appear for
particles in each individual experiment. It is necessary to observe
that the $x $ direction is not special for decay of the
singlet state. Otherwise, additional correlation between spin
projections on the $x $ direction for both particles can take
place. It disturbs cleanliness of the experiment.

If the measurement result for squares of  spin projection on a
picked out axis really depends on the type of device used, then the basic
condition of the no-go theorem of Kochen and Specker appears
outstanding. In that case this theorem cannot be used as an argument
in dispute about the existence of an objective reality.

 \section {The Bell inequality}

We now look how the condition of measurability of a probability
space proves oneself in such important case as deriving the Bell
inequality~\cc {bell1}. There are many variants of this
inequality. We follow the variant proposed in the work~\cc
{chsh}. This variant is usually designated by the abbreviation
CHSH.

Let a particle whose spin is equal to 0 disintegrate into two
particles $A $ and $B $ whose spins are equal to 1/2. These
particles fly apart to a large distance and are registered by the
respective device $D_a $ and $D_b $. The measurements in the
devices are independent. For the particle $A $ the device
$D_a $ measures the spin projection on the $a $ direction, and for
the particle $B $  the device $D_b $ measures the spin projection
on the $b $ direction. We let $ \A $ and $ \B $ denote the
corresponding observables  and let  $A_a $ and $B_b $ denote the
measurement results.

Let us assume that the initial particle possesses the certain
physical reality which can be marked by the parameter $ \lll $. We
shall use the same parameter for the description of physical
realities of products of disintegration. Correspondingly,  it is
possible to consider measurement results of the observables $\A $,
$\B$ as the functions $A_a(\lll)$, $B_b(\lll)$ of the parameter
$\lll$. Let the distribution of the events with respect to the
parameter $\lll$  be characterized by the probabilistic measure $P
(\lll) $: $$ \int dP (\lll) =1, \qquad 0\leq P (\lll) \leq 1. $$

 Let us introduce the correlation function $E (a, b) $:
 \beq {1} E (a, b) = \int dP (\lll) \, A_a (\lll) \, B_b (\lll). \eeq
 Also we  consider the combination
\bea {2}
 I & = & | E (a, b)-E (a, b') | + |E (a', b) +E (a', b') |
 \nn {} & = & \lt| \int dP (\lll) \, A_a (\lll) \, [B_b (\lll) -B
_ {b'} (\lll)] \rt| + \lt| \int dP (\lll) \, A _ {a'} (\lll) \,
[B_b (\lll) +B _ {b '} (\lll)] \rt |. \eea
 For any directions $a $ and $b $
 \beq {3}
 A_a (\lll) = \pm1/2, \quad B_b (\lll) = \pm1/2. \eeq
 Therefore,
\bea {4}
 I & \le &\int dP(\lll)\,[|A_a(\lll)|\,|B_b(\lll)-B_{b'}(\lll)|+ |A_{a'}
 (\lll)| \, |B_b (\lll) +B_{b '} (\lll) |] \nn {} & = &1/2 \int dP (\lll)
\, [|B_b(\lll)-B_{b'}(\lll)|+|B_b(\lll)+B_{b'}(\lll)|].\eea

Due to the equality~\rr {3} for each $ \lll $ one of the
expressions
 \beq{4a}
 |B_b(\lll)B_{b'}(\lll)|, \qquad |B_b(\lll)+B_{b'}(\lll)| \eeq
is equal to zero, and the other is equal to one. Here it is
crucial that the same value of the parameter $ \lll $ appears in
both expressions. Hence, the Bell
inequality (CHSH) is obtained
 \beq {5} I \leq 1/2\int dP(\lll) =1/2. \eeq

The correlation function can be easily calculated within standard
quantum mechanics.  We obtain $$ E (a, b) =-1/4\cos\theta_{ab}, $$
where $ \theta_{ab} $ is the angle between the directions $a $ and
$b $. For the directions $a=0 $, $b =\pi/8 $, $a' = \pi/4 $, $b'
=3\pi/8 $ we have $$ I=1/\sqrt {2}. $$
 It contradicts the inequality~\rr {5}.

The experiments carried out correspond to quantum-mechanical
calculations and do not confirm the Bell inequality. These results
are considered as deciding certificates against the hypothesis
about existence of an objective local reality in quantum physics.
It is easy to make sure that, if to take into account peculiarity
of the application of probability theory to quantum systems,  it
is impossible to carry out such a derivation of the Bell inequality.

Because in a quantum case the $ \sss $-algebra and,
correspondingly, the probabilistic measure depend on the
measuring device used, it is necessary to make replacement $dP(\lll)
\to dP_{ab}(\om)$  in the equation~\rr{1}. Besides, the subset
$\Om_{ab}$ is necessary to take as the regions of  integration
instead of set $ \Om $. Here, $ \Om_{ab} $  is the  set of
physical states which appear in the experiments using the device
for measurement of the spin projections on  the axes $a$ and $b$.
In real experiment (collection of experiments) the sample
$\Om_{ab}$ is finite, in ideal experiment (collection of
experiments) this sample is denumerable. Therefore, we can always
consider that $\Om_{ab}$ is random denumerable sample of the space
$ \Om $.

Thus the equation~\rr{1} should be rewritten in the form
 $$ E (a, b) = \int_{\Om_{ab}} dP_{ab}(\om)\, A_a (\om)\, B_b(\om.) $$
 Now the equation~\rr {2} looks
 \bea {8}
  I & = & \lt | \int_{\Om_{ab}}
dP_{ab}(\om)\,A_a (\om)\, B_b(\om)- \int_{\Om_{ab'}}
dP_{ab'}(\om)\, A_a(\om)\,B_{b'}(\om) \rt | + \nn {} & + & \lt |
\int_{\Om_{a'b}} dP_{a'b}(\om)\,A_{a'} (\om)\,B_b(\om)+
\int_{\Om_{a'b'}} dP_{a'b'}(\om)\,A_{a'}(\om)\, B_{b'}(\om \rt |.
\eea

If the directions $a $ and $a' $ ($b $ and $b' $) are not parallel
to each other, then the observables the $ \A\B $, $ \A\Bb $, $ \Aa\B $, $
\Aa\Bb $ are mutually incompatible. There is no united physically
allowable $ \sss $-algebra which corresponds all these
observables. This implies that there is no united  probabilistic
measure for these observables. Correspondingly, four integrals in
the equation~\rr{8} cannot be united in one.

The sets $\Om_{ab}$, $\Om_{a'b}$, $\Om_{ab'} $, $\Om_{a'b'} $ are
different random samples of the space $ \Om $. The probability of
their intersection is equal to zero because the space $ \Om $ has
a potency of the continuum. Therefore, with the probability one
there is no physical state which could appear in expressions of
type~\rr{4a}. Correspondingly, transition to an inequality of
type~\rr{5} is impossible.

Thus, in the quantum case the hypothesis about existence of a
local objective reality does not drive to the Bell inequalities.
Therefore, the numerous experimental verifications of the Bell
inequalities, which were carried out earlier and are carried out
now,  loose solid theoretical base. Of course, experimental
confirmation of the Bell inequalities would be of outstanding
significance. However, the negative result proves nothing.

\section { Conclusions}

As a rule the experimental investigation in connection with a question of
the existence of local objective reality in quantum physics  is reduced to
examination of an interference pattern. It is possible to get
acquainted with some examples in the paper by Zeilinger~\cc
{zeil} quoted above.

Evidently the interference pattern is determined by the
probabilistic distribution describing physical process under
consideration. At the same time, as shown above, type of
measuring device, which is used in present observation, determines
to a large degree the probabilistic distribution.

In such context, results of Dopfer~\cc {dop} described in
paper~\cc {zeil} look quite natural. He has practically carried
out the experiment similar to gedanken experiment with a so-called
Heisenberg microscope. In this experiment the presence or absence
of an interference pattern depends on what distance from the lens
the detector is placed on.

It then follows that the concept of a local objective reality and,
traced back to Bohr, "the situational (contextual) approach" are
not in such antagonistic contradiction as it is considered to be.
It is quite allowable that there is a physical reality which is
inherent to the quantum object under consideration and which
predetermines the result of any experiment. However, this result can
depend on conditions in which this experiment is carried out. One
of these conditions is the classical characteristic (type) of the
measuring device, which is used in a concrete case.

\begin {thebibliography} {20}

\bb {hom}  D.~Home, M.A.B.~Whitaker ,  Phys. Rep. 210 (1992)  223.
\vspace{-2.5mm}

\bb{kol}  A.N.~Kolmogorov,  Foundation of the theory of
probability, Chelsea, New York (1956).\vspace{-2.5mm}

\bb{nev} J.~Neveu, Bases matematiques du calcul des probabilites,
Masson, Paris (1964). \vspace {-2.5mm}

 \bb{bell1}J.S.~Bell, Physics, 1 (1965) 195. \vspace{-2.5mm}

\bb{bell2} J.S.~Bell, J. Phys.C2 42 (1981) 41. \vspace{-2.5mm}

\bb{epr}  A.~Einstein, B.~Podolsky, N.~Rosen, Phys. Rev. 77 (1935)
777.  \vspace {-2.5mm}

 \bb {slav} D.A.~Slavnov, Theor. Math. Phys.
129  (2001) 87; quant-ph/0101139. \vspace {-2.5mm}

\bb{slav1}D.A.~Slavnov, Theor. Math. Phys. 132  (2002)  1364;
quant-ph/0211053. \vspace {-2.5mm}

\bb{zeil} A.~Zeilinger, Rev. Modern. Phys.  71 (1999) S289.
\vspace {-2.5mm}

\bb{ksp}S.~Kochen, E.P.~Specker,  Journ. of Mathematics and
Mechanics, 17 (1967) 59. \vspace {-2.5mm}

\bb{chsh} J.F.~Clauser, M.A.~Horn,  A.~Shimony, R.A.~Holt,
 Phys. Rev. Lett. 23 (1969) 880.\vspace{-2.5mm}

\bb{dop} Dopfer B. Ph.D. thesis (Universiti of Innsbrug) 1998.

\end{thebibliography}

\end{document}